# STATUS REPORT ON THE
# UNITED NATIONS BASIC SPACE SCIENCE INITIATIVE (UNBSSI):

**Basic Space Science (BSS, 1991-2004), International Heliophysical Year (IHY, 2005-2009), and International Space Weather Initiative (ISWI, 2010-2012)**

Hans J. Haubold and Sharafat Gadimova

Office for Outer Space Affairs, United Nations, Vienna International Centre, A-1400 Vienna, Austria

Abstract. Since 1990, the UN Programme on Space Applications leads the United Nations Basic Space Science Initiative by contributing to the international and regional development of astronomy and space science through annual UN/ESA/NASA/JAXA workshops on basic space science, International Heliophysical Year 2007, and the International Space Weather Initiative. Space weather is the conditions on the Sun and in the solar wind, magnetosphere, ionosphere and thermosphere that can influence the performance and reliability of space-borne and ground-based technological systems and can endanger human life or health. The programme also coordinates the development of IHY/ISWI low-cost, ground-based, world-wide instrument arrays. To date, 14 world-wide instrument arrays comprising approximately 1000 instruments (GPS receivers, magnetometers, spectrometers, particle detectors) are operating in more than 71 countries. The most recent workshop was hosted by the Republic of Korea in 2009 for Asia and the Pacific. Annual workshops on the ISWI have been scheduled to be hosted by Egypt in 2010 for Western Asia, Nigeria in 2011 for Africa, and Ecuador in 2012 for Latin America and the Caribbean.

*The Initiative*

The UBSSI is a long-term effort for the development of space science and regional, and international cooperation in this field on a worldwide basis, particularly in developing nations. A series of workshops on BSS was held from 1991 to 2004 (India 1991, Costa Rica and Colombia 1992, Nigeria 1993, Egypt 1994, Sri Lanka 1995, Germany 1996, Honduras 1997, Jordan 1999, France 2000, Mauritius 2001, Argentina 2002, and China 2004)[1]; and addressed the status of astronomy in Asia and the Pacific, Latin America and the Caribbean, Africa, and Western Asia. One major recommendation that emanated from these workshops was that small astronomical facilities should be established in developing nations for research and education programmes at the university level. Subsequently, material for teaching and observing programmes for small optical telescopes were developed or recommended and astronomical telescope facilities have been inaugurated in a number of nations. Such workshops on BSS emphasized the particular importance of astrophysical data systems and the virtual observatory concept for the development of astronomy on a worldwide basis. Pursuant to resolutions of the United Nations Committee on the Peaceful Uses of Outer Space (UNCOPUOS) and its Scientific and Technical Subcommittee, since 2005, these workshops focused on IHY 2007 (United Arab Emirates 2005,

---

[1] See http://www.seas.columbia.edu/~ah297/un-esa/

India 2006, Japan 2007, Bulgaria 2008, South Korea 2009)[2]. Starting in 2010, the workshops will focus on the ISWI as recommended in a three-year-workplan as part of the deliberations of UNCOPUOS [3]. United Nations/ European Space Agency/ National Aeronautics and Space Administration /Japan Aerospace Exploration Agency workshops on the ISWI have been scheduled to be hosted by Egypt in 2010 for Western Asia, Nigeria in 2011 for Africa, and Ecuador in 2012 for Latin America and the Caribbean..

Already in 2004, the Scientific and Technical Subcommittee of UNCOPUOS agreed that solar-terrestrial physics was important in exploring the solar corona and understanding the functioning of the Sun; understanding the effects that the variability in the Sun can have on the Earth's magnetosphere, environment and climate; exploring the ionized environments of planets; and reaching the limits of the heliosphere and understanding its interaction with interstellar space. The Subcommittee also agreed that, as society became increasingly dependent on space-based systems, it was vital to understand how space weather, caused by solar variability, could affect, among other things, space systems and human space flight, electric power transmission, high-frequency radio communications, global navigation satellite systems (GNSS) signals and long-range radar, as well as the well-being of passengers in high altitude aircraft. In the period of time from 2005 to 2009, UNCOPUOS implemented the IHY 2007, a worldwide campaign to better understand solar-terrestrial interaction.

The IHY 2007 was an international programme of scientific collaboration involving thousands of scientists from all United Nations Member States that was conducted from 2005 to 2009. Along with programmes devoted to research, outreach, and historical preservation of the International Geophysical Year of 1957 (IGY 1957), activities of IHY 2007 included the deployment of new instrumentation arrays especially in developing countries, and an extensive education and public outreach component.

It was recognized early in the planning of IHY 2007 that the understanding of the global ionosphere and its linkage to the near-Earth space environment was limited by the lack of observations in key geographical areas. To address this need, a series of United Nations/ European Space Agency/ National Aeronautics and Space Administration /Japan Aerospace Exploration Agency workshops was held to facilitate collaborations between research scientists in scientifically interesting geographic locations, and researchers in countries with the expertise in building scientific instrumentation. From these workshops, science teams emerged, implementing so-called Coordinated Investigation Programmes (CIPs). Each team consisted of a lead scientist who provided the instruments or fabrication plans for instruments in the array. Support for local scientists, facilities, and data acquisition was provided by the host nation. As a result of the IHY 2007 programme, scientists from many countries continue participating in instrument operation, data collection, analysis, and publication of scientific results. The instrument deployment programme was one of the major successes of the IHY 2007. Arrays of small instruments such as magnetometers to measure Earth's magnetic field, radio antennas to observe solar coronal mass ejections, Global Positioning Systems (GPS) receivers, Very Low Frequency (VLF) radio receivers, and muon particle detectors to observe energetic particles were installed around the world. These arrays continue to provide global measurements of heliospheric phenomena.

---

[2] See http://www.unoosa.org/oosa/SAP/bss/ihy2007/index.html
[3] See http://www.stil.bas.bg/ISWI/

In 2009, UNCOPUOS endorsed a recommendation from its Scientific and Technical Subcommittee to implement ISWI under a three year workplan.

Building on the instrument arrays, and to continue coordinated heliophysics research, in February 2009, the ISWI was proposed as a new agenda item to be addressed by the Scientific and Technical Subcommittee of UNCOPUOS. Through ISWI, coordinated international research will continue on universal processes in the solar system that affect the interplanetary and terrestrial environments, and there will be continued coordination on the deployment and operation of new and existing instrument arrays aimed at understanding and predicting the impacts of space weather on the Earth and the near-Earth environment. The ISWI agenda item was endorsed by UNCOPUOS in June 2009, and by the UN General Assembly in October 2009.

Participation in ISWI is open to scientists from all countries as either instrument hosts or as instrument providers. The ISWI will be governed by a Steering Committee. ISWI will be supported by United Nations (UN), European Space Agency (ESA), National Aeronautics and Space Administration (NASA), Japan Aerospace Exploration Agency (JAXA), and the International Committee on Global Navigation Satellite Systems (ICG)[4].

*Objectives*

The ISWI will help develop the scientific insight necessary to understand the physical relationships inherent in space weather, to reconstruct and forecast near-Earth space weather, and to communicate this knowledge to scientists and to the general public. This will be accomplished, as successfully proven for IHY 2007, by (1) continuing to deploy new instrumentation, (2) developing data analysis processes, (3) developing predictive models using ISWI data from the instrument arrays to improve scientific knowledge and to enable future space weather prediction services, and (4) continuing to promote knowledge of heliophysics through education and public outreach.

*Instrument Array Development*

The ISWI will continue to expand and deploy new and existing instrument arrays following the successful model demonstrated during the IHY 2007. Each instrument team is led by a single scientist. The lead scientist or principle investigator, funded by his/her country, provides instrumentation (or fabrication plans) and data distribution. In a few cases, where resources allow, the host country will pay for the instrument. The host country provides the workforce, facilities, and operational support necessary to operate the instrument. This would typically be at a local university or government laboratory. Host scientists become part of the science team. All data and data analysis activities are shared within the science team, and all scientists participate in publications and scientific meetings where possible. Through workshops and other means, the ISWI will actively seek to identify additional instruments and instrument providers that could benefit from the ISWI process, as well as new instrument hosts.

---

[4] See http://www.icgsecretariat.org

*Data Coordination and Analysis*

The ISWI programme will promote the coordination of data products in a form useful for input into physical models of heliospheric processes. These data will be used for both retrospective analysis aimed at physical understanding of space weather, and for predictive models to predict future space weather conditions. To be useful for space weather prediction, data must be available in near real-time. However, today internet connections are intermittent or slow in many locations in the developing world, making near real-time data return impossible. Eventually, as internet connectivity improves, these data will be made available in near real-time in a form, which can be incorporated into predictive models. In the near term, other strategies like data transfer during selected time periods, or on recorded media like DVDs and tapes, will be adequate for the retrospective scientific studies of space weather events, and the development of physical models. Data from the instrument arrays will be deposited in publicly available archives. For the most part, these will be existing data archives, like the virtual observatory systems which are currently under development. This will make data from ISWI instruments available to the broader community of researchers. To improve the coordination of the data and to enhance their value for future real-time prediction services, planning will begin for the availability and the interoperability of these data. Although the infrastructure and the institutional resources may not yet exist in many locations to support the real-time dissemination of quality-controlled data, it is important to begin the discussion now of data standards and the expectation of continuous operation so that the development of data systems and the discussions of future resource allocations can be done with this goal in mind.

*Training, Education, and Outreach*

During the IHY 2007, space science schools in a number of countries provided related training to hundreds of graduate students and new researchers. The ISWI will continue to provide support for space science schools. The ISWI will continue to promote space science and the inclusion of space science curricula in universities and graduate schools. This has been most effective when combined with the installation of instrumentation at the respective universities. The ISWI will continue to support public outreach projects. It is essential to communicate the excitement and the relevance of heliophysical research to scientists from other disciplines, and to the public at large. Through ISWI, public outreach materials unique to the ISWI will continue to be developed, and its distribution will be coordinated through individual contacts and outreach workshops.

*Monitoring Solar-terrestrial Interaction at the United Nations Office at Vienna*

Earth's ionosphere reacts strongly to the intense X-ray and ultraviolet radiation released by the Sun during solar events. Stanford's Solar Center developed inexpensive space weather monitors that scholars around the world can use to track changes to the Earth's ionosphere. Two versions of the monitors exist – a low-cost version named Sudden Ionospheric Disturbances (SID) designed to detect solar flares; and a more sensitive version named Atmospheric Weather Electromagnetic System of Observation, Modeling, and Education (AWESOME) that provides both solar and nighttime research-quality data. Through UNBSSI, such monitors have been deployed to high schools and universities in developing nations of the world for the ISWI. The monitors come preassembled, the hosts build their own antenna, and provide a PC to record the data and an internet connection to share their data with worldwide network of SIDs and AWESOMEs. These networks are advancing the understanding of the fundamental heliophysical processes that govern the Sun, Earth and heliosphere, particularly phenomena of space weather. Monitoring the fundamental processes responsible for solar-terrestrial coupling are vital to being able to understand the influence of the Sun on the near-Earth environment. A SID monitor is successfully operating at the United Nations Office at Vienna (UNOV) and will be extended to an AWESOME shortly. This project will also be supported by the programme on GNSS applications, implemented through the ICG.

*Further Reading*

W. Wamsteker, R. Albrecht, and H.J. Haubold: Developing Basic Space Science World-Wide: A Decade of UN/ESA Workshops, Kluwer Academic Publishers, Dordrecht 2004.

B.J. Thompson, N. Gopalswamy, J.M. Davila, and H.J. Haubold: Putting the "I" in IHY: The United Nations Report for the International heliophysical Year 2007, Springer, Wien and New York 2009.

Organized by the United Nations, the European Space Agency (ESA), the National Aeronautics and Space Administration (NASA) of the United States of America and the Japan Aerospace Exploration Agency (JAXA), the fifth UN/ESA/NASA/JAXA Workshop on basic space science and the International Heliophysical Year 2007, was hosted by the Korean Astronomy and Space Science Institute (KASI). The four previous workshops in this series of activities were hosted by the Governments of the United Arab Emirates, in 2005, of India, in 2006, of Japan, in 2007, and of Bulgaria, in 2008 (UN documents A/AC.105/856, A/AC.105/489, A/AC.105/902, A/AC.105/919, respectively). These workshops were a continuation of the series of workshops on basic space science that were held between 1991 and 2004 and that were hosted by the Governments of India (A/AC.105/489), Costa Rica and Colombia (A/AC.105/530), Nigeria (A/AC.105/560/Add.1), Egypt (A/AC.105/580), Sri Lanka (A/AC.105/640), Germany (A/AC.105/657), Honduras (A/AC.105/682), Jordan (A/AC.105/723), France (A/AC.105/742), Mauritius (A/AC.105/766), Argentina (A/AC.105/784) and China (A/AC.105/829).

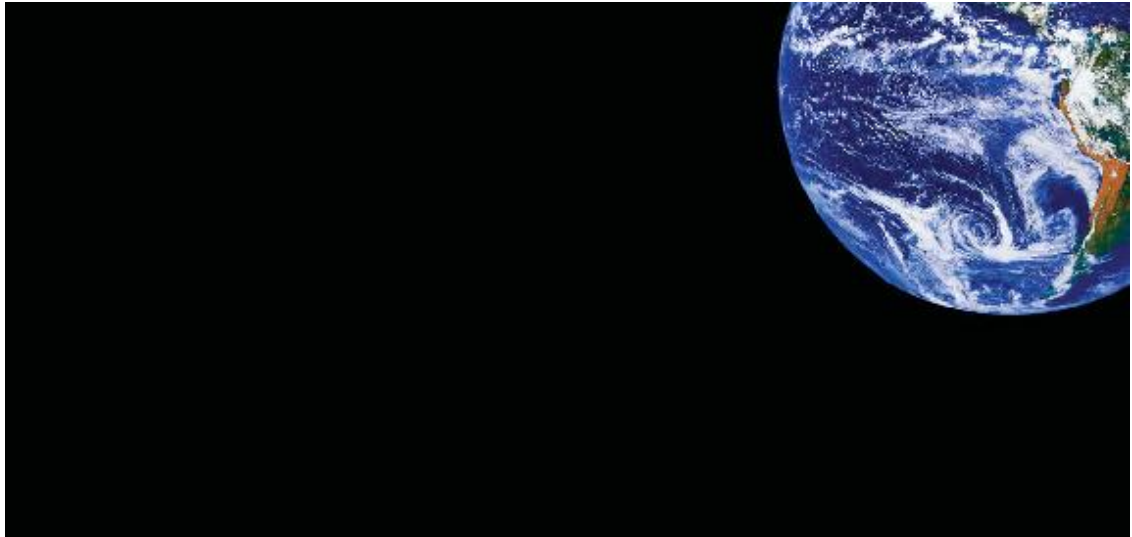
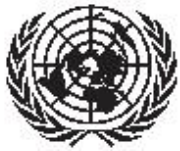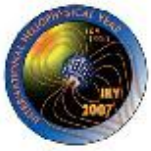
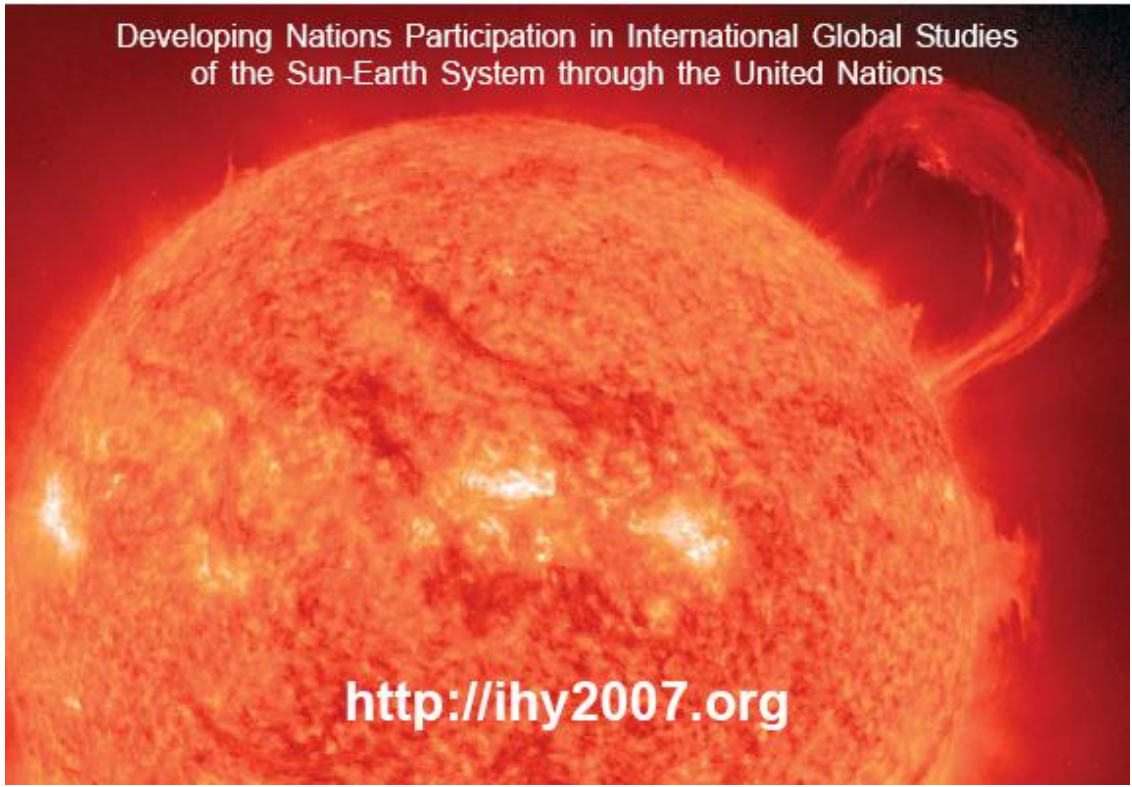

*Credit: UNOOSA*

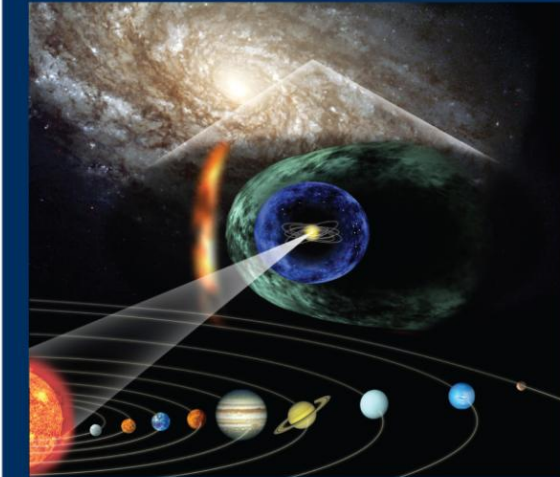
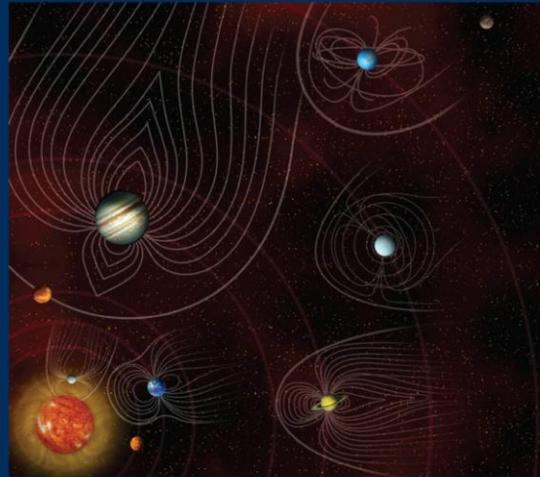
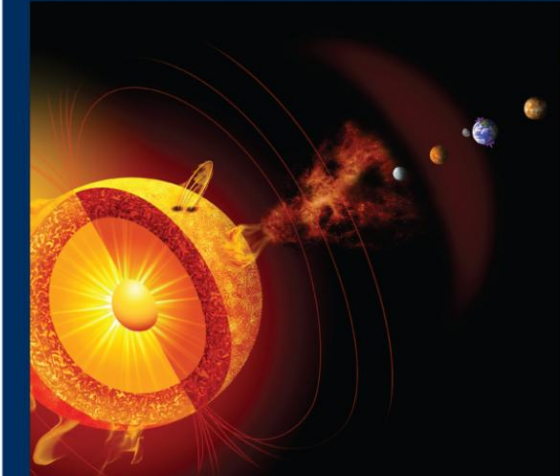
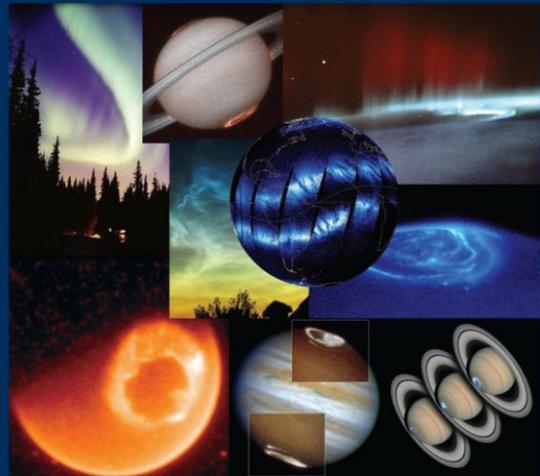

*Credit: NASA*